\documentclass[aps,pre]{revtex4}

\usepackage{epsfig}
\usepackage{graphicx}
\usepackage[dvips]{color}

\def\be{\begin{equation}}
\def\ee{\end{equation}}
\def\ba{\begin{array}}
\def\ea{\end{array}}
\def\bea{\begin{eqnarray}}
\def\eea{\end{eqnarray}}

\newcommand{\rev}[1]{{\textcolor[named]{Black}{#1}}}

\begin{document}

\title{Path independent integrals to identify
localized plastic events in two dimensions}

\author{Mehdi Talamali$^1$, Viljo Pet\"aj\"a$^1$, Damien Vandembroucq$^1,2$
 and St\'ephane Roux$^3$
}

\affiliation
{1: Unit\'e Mixte CNRS/Saint-Gobain ``Surface du Verre et Interfaces''\\
39 Quai Lucien Lefranc, 93303 Aubervilliers cedex, FRANCE\\
2: Laboratoire PMMH, ESPCI/CNRS/Paris 6/Paris 7\\
10 rue Vauquelin, 75231 Paris cedex 05, France\\
3: LMT-Cachan\\
ENS de Cachan / CNRS-UMR 8535 / Universit\'{e} Paris 6/PRES UniverSud\\
61 avenue du Pr\'{e}sident Wilson, F-94235 Cachan Cedex, France.}

\begin{abstract}
We use a power expansion representation of plane elasticity
complex potentials due to Kolossov and Muskhelishvili, to compute
the elastic fields induced by a localized plastic deformation
event. Far from its center, the dominant contributions correspond
to first order singularities of quadrupolar and dipolar symmetry
which can be associated respectively to pure deviatoric and pure
volumetric plastic strain of an equivalent circular inclusion.
Constructing holomorphic functions from the displacement field
and its derivatives, it is possible to define path independent
Cauchy integrals which capture the amplitudes of these
singularities. Analytical expressions and numerical tests on
simple finite element data are presented. The development of such
numerical tools is of direct interest for the identification of
local structural reorganizations which are believed to be the key
mechanisms for plasticity of amorphous materials.
\end{abstract}

\maketitle

\section{Introduction}
\label{intro}

Plasticity of amorphous materials has motivated an increasing
amount of studies in recent years. In the absence of underlying
crystalline lattice in materials such as foams, suspensions or
structural glasses, it is generally accepted that plastic
deformation results from a succession of localized structural
reorganizations\cite{BulatovArgon94a,BulatovArgon94b,
BulatovArgon94c,FalkLanger-PRE98}. Such changes of local
structure release part of the elastic strain to reach a more
favorable conformation and induce long range elastic fields. The
details of such local rearrangements and of the internal stress
they induce obviously depend on the precise structure of the
material under study, and its local configuration . However, the
important observation is that outside the zone of reorganization,
{\em a linear elastic behavior prevails}.  Therefore, elastic
stresses can be decomposed onto a multipolar basis and
independently of the material details, it is possible to extract
singular, scale-free, dominant terms which can be associated to a
global pure deviatoric or pure volumetric local transformations
of an equivalent circular inclusion. In particular, the elastic
shear stress induced by a localized plastic shear exhibits a
quadrupolar symmetry. This observation has motivated the
development of statistical models of amorphous plasticity at
mesoscopic scale based upon the interaction of disorder and long
range elastic
interactions\cite{BVR-PRL02,Picard-PRE04,Lemaitre-preprint06,Jagla-PRE07}.
In the same spirit statistical models were also recently
developed to describe the plasticity of poly-crystalline
materials\cite{Zaiser-JSM05}.  Several numerical studies have
been performed recently to identify these elementary localized
plastic events in athermal or molecular dynamics simulations of
model amorphous material under shear
\cite{Tanguy-EPJE06,Maloney-PRE06}.

The question remains how to identify and analyze these transformation
zones.  In analogy with the path independent Rice
J-integral\cite{Rice-JAM68} developed to estimate the stress intensity
factor associated to a crack tip stress singularity, we aim here at
capturing the stress singularity induced by the local plastic
transformation which can be treated as an Eshelby
inclusion\cite{Eshelby57}. In two dimensions, we develop a simple
approach based upon the Kolossov-Muskhelishvili (K\&M) formalism of
plane elasticity\cite{Muskhelishvili}. This is an appealing
pathway to the solution since these zones will appear as poles for the
potentials, and hence Cauchy integrals may easily lead to contour
integral formulation which are independent of the precise contour
geometry, but rather rely on its topology with respect to the
different poles which are present.

Although these techniques have been mostly used in the context of
numerical simulations in order to estimate stress intensity
factors from finite element simulations, they are now called for
to estimate stress intensity factors from experimentally measured
displacement field from e.g. digital image correlation
techniques.  In this case, interaction integral
techniques\cite{Rethore-IJF05} or least square
regression\cite{Roux-IJF06} techniques have been applied.  Noise
robust variants have also been proposed\cite{Rethore-EFM07}.
These routes could also be followed in the present case.

 Though the present work is
restricted to two dimensions due to the complex potential
formulation, similar questions can be addressed for the three
dimensional version of this problem using the same strategy but a
different methodology. In the following, we briefly recall the
K\&M formalism, and we give analytic expressions of contour
integrals allowing to capture the singular elastic fields and we
present a few numerical results based on a finite element
simulation supporting our analytical developments.

\rev{In Section 2, we present the theoretical basis of our
approach in terms of singular elastic fields, while Section 3
introduce the contour integral formulation.  In section 4, a
numerical implementation based on finite element simulations is
presented together with the results of the present approach. This
application allows to evaluate the performance and limitations of
the contour integral procedure and check the detrimental effect
of discreteness. Section 5 presents the main conclusions of our
study.}

\section{Potential formulation}

In two dimensions the Kolossov and Muskhelishvili potentials
(K\&M) can be used to write the elastic stress and displacement
fields ${\bf U}$ and $\sigma$\cite{Muskhelishvili}. Using a
complex formulation, we introduce the elastic displacement
${\mathbf U}=U_x +i U_y$ and the stress tensor field through two
functions, the real trace $S_0=\sigma_{xx}+\sigma_{yy}$ and the
complex function ${\mathbf S} = \sigma_{yy}- \sigma_{xx}+
2i\sigma_{xy}$. \rev{In the framework of linear elasticity,
balance and compatibility equations can be rewritten as:}
\bea
S_{0,z}- {\mathbf S}_{,\overline{z}} 
=0\;,\label{balance}\\
S_{0,z\overline{z}} 
=0 \label{compatibility}\;,
\eea
where $z=x+iy$ is the complex coordinate and the notation
$A_{,x}$ is used to represent the partial derivative of field $A$
with respect to coordinate $x$. \rev{Note that we assumed zero
surface density force and that equation (\ref{compatibility}) is
here the classical Beltrami equation which expresses the
kinematic compatibility condition in terms of stress}.  The
general solution to these equations can be obtained through the
introduction of two holomorphic functions $\varphi$ and $\psi$,
called the K\&M potentials.  The displacement and the stress
field can be written \cite{Muskhelishvili}

\bea
2\mu {\mathbf U}=&\kappa \varphi(z) -z\overline{\varphi'(z)}
 -\overline{\psi(z)}\;,\\
S_0
=&2 \left[ \varphi'(z)+\overline{\varphi'(z)} \right]\;,\\
{\bf S}
=&2 \left[ \overline{z}\varphi''(z)+\psi'(z) \right]\;,
\eea
where $\mu$ is the elastic shear modulus and $\kappa=(3-4\nu)$
for plane strain and $\kappa=(3-\nu)/(1+\nu)$ for plane stress,
$\nu$ being the Poisson's ratio.

\section{Plastic inclusion and singularity approach in 2D}

\subsection{Singular terms associated to plastic inclusion}

This K\&M formalism can be applied to two-dimensional inclusion
problems\cite{Jawson-PCPS60,Mathiesen-preprint07}.  Let us consider
the case of a small inclusion of area ${\cal A}$ experiencing plastic
deformation and located at the origin of the coordinate system $z=0$.
It is assumed that the stress is a constant at infinity.  Outside the
inclusion, the K\&M potentials can be expanded as a Laurent series as
\bea
\varphi(z)= \alpha^{out} z +\sum_{n=1}^{\infty}\frac{ \varphi_n}{z^n} \;,\quad
&\displaystyle \psi(z)= \beta^{out} z +\sum_{n=1}^{\infty}\frac{ \psi_n}{z^n}
\label{Laurent}
\eea
The linear terms can be easily identified as corresponding to uniform
stresses while constant terms (omitted here) would lead to a rigid
translation. It can be shown in addition that the dominant singular
terms $\varphi_1/z$ and $\psi_1/z$ can be associated to the elastic
stress induced by the plastic deviatoric and volumetric of an
equivalent circular inclusion of area ${\cal A}$. Namely, considering
a circular inclusion experiencing a plastic shear strain $\gamma_p$
and a plastic volumetric strain $\delta_p$ we
have\cite{Jawson-PCPS60}:
\be
\varphi_1 =\frac{2i\mu {\cal A}\gamma_p}{\pi(\kappa+1)}\;,\quad
\psi_1 = -\frac{2\mu {\cal A} \delta_p}{\pi(\kappa+1)} \;.
\label{phi1-psi1}
\ee
 In particular, for a pure shear plastic event
we obtain a quadrupolar symmetry:
\be \sigma_{xy}= -\frac{2\gamma_p \mu}{\kappa+1}\frac{\cal A}{\pi r^2}
\cos(4\theta) \ee
Note that we have in general to consider a complex value of
$\gamma_p$ to include the angular dependence of the principal
axis.  In contrast, the amplitude $\psi_1$ is a real number (note
that the imaginary part would correspond to a point-like torque
applied at the origin).

\subsection{Generic character of the expansion}

Because of the well-known property of Eshelby circular inclusion,
the above expansion limited to the $\psi_1$ and $\phi_1$ terms
only is the exact (outer) solution of a uniform plastic strain
distributed in the inclusion, and vanishing stress at infinity.
However, one should note that this result is much more general.
Indeed, it is seen that the physical size of the inclusion does
not enter into the solution but through the products ${\cal
A}\gamma_p$ and ${\cal A}\delta_p$.  Therefore, a smaller
inclusion having a larger plastic strain may give rise to the
very same field, provided the products remain constant.
Therefore, one can consider the prolongation of the solution to a
point-like inclusion (with a diverging plastic strain), as being
equivalent to the inclusion.

Then from the superposition property of linear elasticity, a
heterogeneous distribution of plastic strain $\gamma_p(x)$ in a
compact domain, ${\cal D}$, will give rise to such a singularity
with an amplitude equal to
 \be
 [{\cal A}\gamma_p]_{eq}=\int\!\!\!\int_{\cal D} \gamma_p(x) dx
 \ee
and the same property would hold separately for the volumetric
part.  As a particular case, one finds a uniform plastic strain
for an inclusion of arbitrary shape.

This is the key property that allows to capture the equivalent
plastic strain of an arbitrary complex configuration, for the
above mentioned application to amorphous media.  In fact, this is
even the only proper way of defining the plastic strain for a
discrete medium such as encountered in molecular dynamics
simulations.    The far-field behavior of the displacement and
stress field can be accurately modelled, and without ambiguity by
a continuum approach, and thus the above result will hold.  In
contrast, locally, the large scale displacement of several atoms
may render difficult the direct computation of the equivalent
plastic strain experienced in such an elementary plastic event.

Let us however stress one difficulty: As the above argument
ignores the details of the action taking place within the
``inclusion'', plasticity has to be postulated.  However, a
damaged inclusion, where the elastic moduli have been softened by
some mechanism, or even a non-linear elastic inclusion at one
level of loading would behave in a similar way as the above
plastic inclusion.  Obviously, to detect the most relevant
physical description, one should have additional informations,
say about unloading.  If the above amplitudes remain constant
during unloading, plasticity would appear appropriate.  If the
amplitude decreases linearly with the loading, then damage is
more suited.  Finally, if the amplitudes varies reversibly with
the loading, non-linear elasticity might be the best description.
Thus, although one should be cautious in the interpretation,
local damage detection from the far field may also be be tackled
with the same tools.

\subsection{Path independent contour integrals}

In two-dimensions, this multipole expansion formalism in the
complex plane suggests to resort to contour integrals to extract
the singularities. However the displacement field is not a
holomorphic function and cannot be used directly for that
purpose. The strategy of identification of the singularities
$\varphi_n$ and $\psi_n$ thus consists of expressing the
potentials from the displacement field and its derivatives in
order to extract the singularities {\it via} Cauchy integrals. We
now simply express the displacement field and its derivatives:
\bea
2\mu U&=&\kappa \varphi(z)
-z \overline{\varphi'(z)}
- \overline{\psi(z)}\;,\\
2\mu U_{,z}&=& \kappa\varphi'(z)-\overline{\varphi'(z)}
\;,\\
2\mu U_{,\overline{z}}&=& -z \overline{\varphi''(z)}
  -\overline{\psi'(z)}\;,\\
2\mu U_{,z\overline{z}}&=& -\overline{\varphi''(z)}\;. \eea
This gives immediately:
  \bea
\varphi'(z)&=&\frac{2\mu}{\kappa -1} \left[ \kappa U_{,z}+\overline{U_{,z}} \right]\;,\\
  \varphi''(z)
  &=&-2\mu \overline{U_{,z\overline{z}}}
  =-\frac{\mu}{2}\overline{\nabla^2 U}\;,\\
  \psi'(z)
&=&  -2\mu \left[ \overline{U_{,\overline{z}}}
  -\frac{\overline{z}}{4}\overline{\nabla^2 U}\right]\;. \eea
Note that, except a multiplicative constant, the two last expressions
are independent of materials properties. In light of the expansion
(\ref{Laurent}) of $\varphi$ and $\psi$ in Laurent series, if an
anti-clockwise contour integration is considered along a path ${\cal C
}$, Cauchy residues can be formed as
  \bea
  \label{eq:phi-psi}
  \varphi_n &= &\frac{i\mu}{4\pi n(n+1)}
  \int_{\cal C} z^{n+1}\overline{\nabla^2 U} dz\;,\\
  \psi_n &= &\frac{i\mu}{\pi n}\int_{\cal C} z^n
  \left[  \overline{U_{,\overline{z}}}
  - \frac{1}{4}\overline{z} \overline{\nabla^2 U} \right] dz\;.
  \eea
 These expressions can thus be obtained from the sole
knowledge of the displacement field, a quantity which can be
accessed from experiments, or from atomistic simulations. In the
case of first order singularities (see Eq.(\ref{phi1-psi1})), the
residue term thus only depends on the local plastic deformation
(size and amplitude of deformation) and on the Poisson's ratio
$\nu$ of the material.

Reverting to Cartesian coordinates, where the contour is
expressed as a function of the curvilinear abscissa $s$ as
$(x(s),y(s))$, the above expression can be written
 \be\ba{lll}
 \varphi_1 &= &\frac{\mu}{8\pi }
 \int_{\cal C} [-2(xy)+i(x^2-y^2)][(U_{x,xx}+U_{x,yy})-i(U_{y,xx}+U_{y,yy})]
 [\frac{dx}{ds} +i\frac{dy}{ds}]ds\;,\\
 \psi_1 &= &\frac{\mu}{4\pi}\int_{\cal C} \left[
 2 [ix-y][(U_{x,x}-U_{y,y})-i(U_{y,x}+U_{x,y})]\right.\\
 &&\left.- [i(x^2+y^2)]
 [U_{x,xx}+U_{x,yy}-i(U_{y,xx}+U_{y,yy})]
 \right] [\frac{dx}{ds} +i\frac{dy}{ds}]
 ds\;.
 \ea\ee
%

\section{Numerical implementation}

The ultimate goal of a such a method would be to analyze
numerical results obtained from molecular dynamics simulations of
amorphous plasticity where such local structural reorganizations
are expected to take place. This obviously raises the question of
a well defined method for writing the continuous displacement
field from the data of the discrete displacements of particles
\cite{Goldenberg-EPJE02} and more generally the question of the
sensitivity to noise of the above expressions. The first point is
beyond the scope of the present work and we leave it for later
studies. We thus focus on the more restricted question of the
numerical implementation and its efficiency in the case of
artificially noise-corrupted displacement data.

The method relies on contour integrations of derived fields of
the displacement. The latter point induces {\it a priori} a
strong sensitivity to noise. To limit such effects, first and
second derivatives are extracted {\it via} an interpolation of
the local displacement field by polynomial functions of the
spatial coordinates. Moreover, the path independence of the
contour integrals allows to perform spatial averages. We explore
in the following the efficiency of this method on noisy data.

\begin{figure}
\begin{center}
\includegraphics[width=10cm]{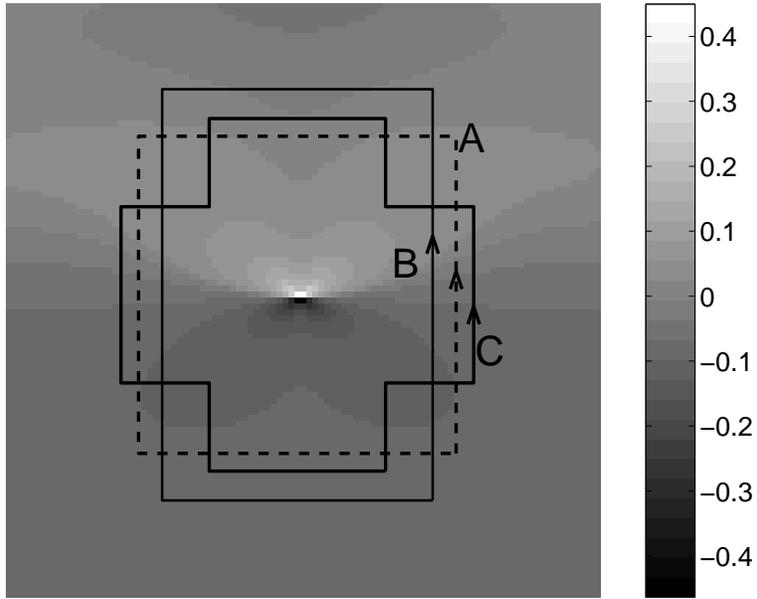}
\end{center}
\caption{\label{Uy-shear100} Map of the displacement field $U_y$
induced by a plastic shear strain of a central square element.
The oriented paths indicate contours for integration.  }
\end{figure}

\subsection{Numerical generation of elastic fields induced by plastic inclusions}

Displacement fields are computed numerically using a finite
element code, with square elements and bi-linear shape functions
$\{1,x,y,xy\}$.  Plane stress conditions of two-dimensional
elasticity are used. The domain is a 150$\times$150 square.
Stress free conditions are enforced all along the domain
boundary.  The Poisson's ratio of the material is $\nu=0.20$.
Since no quantitative values of the stress are used, the value of
the Young's modulus is immaterial.

A plastic strain is implemented at the scale of one single
isolated element.  \rev{Within this element, the strain is the
sum of a plastic uniform strain chosen at will, and an elastic
strain. The latter is computed by solving for the two-dimensional
elastic problem, insuring force balance and kinematic continuity
at all nodes including the nodes of the plastic element.}  The
chosen kinematics is too crude to solve at the scale of one
single element the elastic problem with a good accuracy. However,
remote from the inclusion, the elastic perturbation is well
accounted for, and since all our computations are based on paths
lying at a distance from the inclusion, the formalism should be
applicable. A single element allows us to have a maximum ratio
between inclusion and domain size. The price to pay for this
crude local description is that the quantitative estimate of
${\cal A}\gamma_p$ and ${\cal A}\delta_p$ will differ slightly
from the theoretical expectation.  Nevertheless the path
independence, (size, shape, center, ...) is expected to hold.

We limited ourselves to such a description in order to mimic the
difficulties one may face when having to deal with discrete
element simulations.  Indeed, the chosen finite element shape
functions do not allow us to use this description strictly
speaking in order to compute second order differential operators
on the displacement, since the gradients of the latter are not
continuous across element boundaries.  Therefore, a
regularization will be called for, as detailed below.

\rev{The choice of a regular square lattice is obviously
oversimplified compared with the case of the random lattices
associated with atomistic simulations. However, as $x$ and $y$
directions are obviously equivalent for square elements and as
linearity is preserved by the finite element formulation, this
formulation should not introduce any breaking of symmetry. More
specifically, the displacement field induced by a quadrupole of
principal direction off axis can be obtained by a linear
superposition of $x$ and $y$ components of the displacement field
induced by a quadrupole aligned with the axis weighted by the
sine and cosine of the quadrupole angle.}

Finally,
the finite size of the system is also a specific difficulty
encountered in practice, whereas the above argument uses the
assumption of an infinite domain.  However, such a boundary
condition should not induce additional poles within the domain,
and can be considered as a common practical difficulty
encountered for all practical use of this tool. All these
arguments are possible causes of deviation from the theoretical
expectation of path independence, and it thus motivates a
detailed study of the method stability, robustness and accuracy.

Two test cases are studied. I: a central inclusion experiencing a
shear strain $\gamma_p=1$ (because of linearity, the actual
amplitude is meaningless) along the $x-$axis; II: a central
inclusion experiencing a volumetric contraction $\delta_p=-1$. A
map of the displacement fields $U_y$ in case I is given on Fig.
\ref{Uy-shear100}.

\begin{figure}
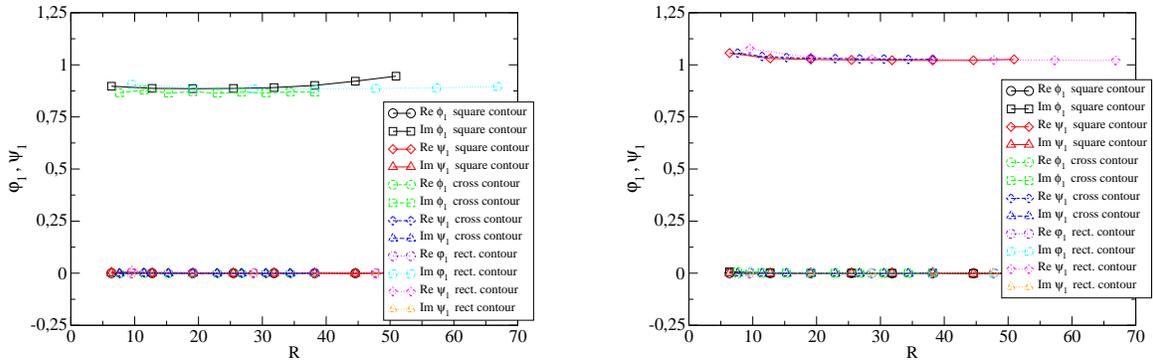

\begin{center}
\includegraphics[width=7cm,angle=0]{TPRV-fig2a.eps}
\hspace{1cm}
\includegraphics[width=7cm,angle=0]{TPRV-fig2b.eps}
\end{center}
\caption{\label{phi-psi-vstaille} Normalized values of numerical
estimates $\varphi_1$ and $\psi_1$ obtained for three families of
contours, respectively square, cross and rectangle shaped and of varying
length $L$ in the case of a displacement field induced by the
plastic shear strain (left) or contraction (right) experienced by
the central element of a square lattice. Theoretical expectations
are $\varphi_1=i\;,\psi_1=0$ (left) and $\varphi_1=0\;,\psi_1=1$
(right).}
\end{figure}

\subsection{Interpolating displacement data}

The key ingredient is to go from a continuous but non-differentiable
displacement field obtained from the finite element simulation to an
evaluation of the second derivative at any point in the domain.

The results which are presented below have been obtained using the
following procedure. A quadratic fit is performed on a square centered
on one node to extract the first and second order derivatives, the
obtained values are used to compute the integrals by quadrature.  An
alternative method has been tested: for an integration from $(x,y)$ to
$(x+1,y)$, a simple fit is performed of the 12 nodes ranging from
$(x-1)$ to $(x+2)$, and from $(y-1)$ to $(y+1)$, by the tensor product
of polynomials $(1,x,x^2,x^3)$ and $(1,y,y^2)$ (12 functions). Then
the integral of all required quantities can be computed.  Estimates of
derivatives using of Fourier Series with and without low pass
filtering have also been performed. All methods give similar results
provided that the area of the region used for interpolation (or
filtering) is comparable.

\subsection{Path independence}

We first check the path independence of the contour integral in the
cases I and II of isolated inclusions. For that purpose, we use two
families of contours, square (A) and cross (B) and rectangular (C) shaped respectively as shown on Fig. \ref{Uy-shear100}. The
size of these contours as well as their center can be varied. Figure
\ref{phi-psi-vstaille} gives a summary of the results. For the three
kinds of contours, we show the real and imaginary parts of the
residues corresponding to Eq. (\ref{eq:phi-psi}). Note that the
numerical results have been normalized according to the theoretical
expectations (\ref{phi1-psi1}) so that the expected numerical values
are $\varphi_1=i\;,\psi_1=0$ in case I (Fig. \ref{phi-psi-vstaille}
left) and $\varphi_1=0\;,\psi_1=1$ in case II
(Fig. \ref{phi-psi-vstaille} right).

These numerical results can be considered as rather satisfactory in
terms of orientation and orthogonality between modes $\varphi_1$ and
$\psi_1$: the measured values of quantities whose expected value is
zero remain typically below $10^{-2}$. When compared to their
theoretical values, $\varphi_1^{\mathrm shear}$ and $\psi_1^{\mathrm
contraction}$ exhibit relative differences of around 5-10\% .  Small
fluctuations (below 5\%) can be found when changing shape and size of
the contours. We already commented on the fact that the finite element
simulation are performed with a single element for the inclusion, a
procedure which is obviously not reliable in terms of accuracy, but
which allows us to have a large ratio between element and system size.

\begin{figure}
\begin{center}
\includegraphics[width=8cm]{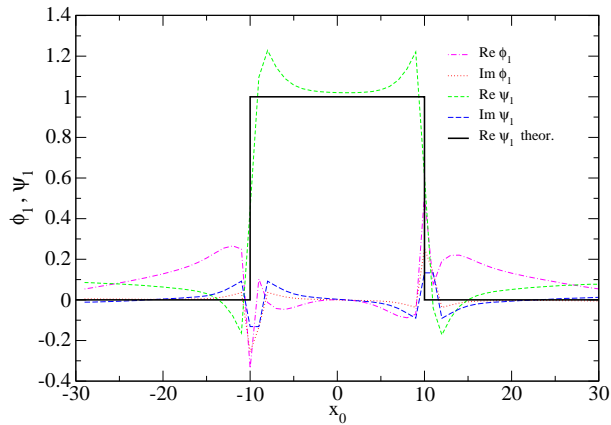}
\end{center}
\caption{\label{phi-vscentre} Normalized values of numerical estimates
$\varphi_1$ obtained for square contours of center $(x_0,0)$ and of
size $M=20$. The expected behavior of $\Re \psi_1$ (unity when the
inclusion lies within the contour, zero otherwise) is represented by
the bold line. Other quantities are expected to be zero.}
\end{figure}

Another test of the dependence of the numerical procedure is on
the sole topology, {\it i.e.} location of the inclusion inside or
outside the contour, we show in addition the dependence of the
measured values of $\varphi_1$ and $\psi_1$ on the location of
the contour center. Fig. \ref{phi-vscentre} shows the singularity
$\psi_1$ measured from the integration of displacement field II
along a square contour centered along the $x$-axis.  Results are
normalized so that the expected value of $\Re \psi_1$ be unity
when the inclusion is within the contour and zero elsewhere.The
contour size is $M=20$. We obtain the expected behavior: the
values of $\Re \psi_1$ shifts form zero to unity depending on the
inclusion is within or outside the contour. Significant
fluctuations (10-20\%) are however observed  when the inclusion
lies in the vicinity of the contour.

\begin{figure}
\begin{center}
\includegraphics[width=8cm]{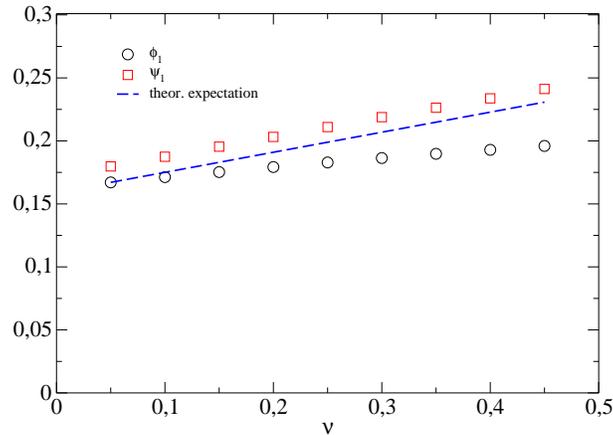}
\end{center}
\caption{\label{phi-psivsnu} \rev{Numerical estimates of
$|\varphi_1|$ and $\psi_1$ obtained for a central inclusion
experiencing a unit shear and a unit contraction respectively as
a function of the Poisson ratio $\nu$. The numerical results
obtained in plane stress conditions for a system of size
$100\times100$ with stress free boundary conditions are compared
with the theoretical expectation $(1+\nu)/2\pi$.}}
\end{figure}

\rev{Finally we test the dependence of the numerical method on
the material properties. In the determination of $\varphi_1$ and
$\psi_1$ (\ref{eq:phi-psi}) as residues, the need to resort to a
second order derivative of the displacement field is balanced by
the fact that the computation can be performed without any
knowledge of the elastic properties of the material. The
independence of the numerical procedure with the Young modulus is
trivially obtained due to the linearity of the FEM computation.
In Fig. \ref{phi-psivsnu} we show the dependence of the numerical
results on the Poisson ratio. FEM computations have been
performed on systems of size $100\times100$ with stress free
boundary conditions and a central inclusion experiencing a unit
shear and a unit contraction respectively. Poisson ratios have
been varied from 0.05 to 0.45 by step of 0.05. The results shown
in the figure compare the numerical estimates obtained for a
square contour of size 40 centered on the inclusion with the
theoretical expectation $\psi_1=\varphi_1=(1+\nu)/2\pi$. The
numerical results show that the volumetric strain is weakly
dependent on Poisson's ratio, but the elementary shear is more
poorly estimated for high Poisson's ratio.  
}

\section{Discussion/conclusion}

The proposed approach is based on a an exact result and hence
theoretically establish a parallel with other types of elastic
singularities (in particular for stress intensity factors which
characterize crack loadings) where similar path integral are well
known. When tested on direct numerical simulations, we could
recover the main topological properties expected in this context:
path independence and detection of  the absence/presence of a
singularity within the contour. However, the quantitative results
proved more disappointing: the method is rather unprecise on the
determination of the prefactor of the singularity and is more
generally rather sensitive to noise.  The main cause is
presumably due to the inconsistent regularity of the displacement
field solution (simple continuity) with the need to resort to
estimates of first and second order differentials.  A piecewise
high order polynomial interpolation is operational for integrals
over finite segments, however, from one segment to the next,
first and second order differentials will display a discontinuous
character, which obviously affect the method and result.
Moreover, being a path integral, the method does not take
advantage of the knowledge of the displacement field at all
points of a domain. To make the method more robust with respect
to noise, different approaches can be followed.  One natural way
is to average the result over different contours, thus
transforming the contour integral into a domain integral.  An
arbitrary weight average can also be considered, and hence, one
could optimize the weight in order to achieve the least noise
sensitivity.  Such a method was explored successfully for cracks
in ref.~\cite{Rethore-EFM07}. Note finally that extensions to
three dimensions obviously require a different technique than
Kolossov and Muskhelishvili potentials, and contour integrals,
however still a linear extraction operator acting on the
displacement field can be computed to provide similarly the
equivalent plastic strain.

\begin{acknowledgements}
V. Pet\"aj\"a acknowledges the financial support of ANR program
PlastiGlass NT05-4\_41640 and of the academy of Finland.
\end{acknowledgements}

\bibliography{plasticity,vdb}   

%
%


\end{document}